\documentclass[journal=apchd5,manuscript=article]{achemso}

\usepackage{chemformula} 
\usepackage[T1]{fontenc} 

\newcommand*{\citen}[1]{%
  \begingroup
    \romannumeral-`\x 
    \setcitestyle{numbers}%
    \cite{#1}%
  \endgroup   
}
\author{Andres Forrer}
\email{aforrer@phys.ethz.ch}
\author{Martin Francki\'e}
\author{David Stark}
\author{Tudor Olariu}
\author{Mattias Beck}
\author{J\'er\^ome Faist}
\author{Giacomo Scalari}
\email{scalari@phys.ethz.ch}

\affiliation[ETH all]{Institute of Quantum Electronics, Eidgen\"ossische  Technische Hochschule Z\"urich, Switzerland}

\title[Manuscript Title]
  {Photon-driven broadband emission and RF injection locking of THz quantum cascade laser frequency combs}

\abbreviations{QCL, FC, THz}

\begin{document}
\begin{tocentry}
\includegraphics[width=0.8\linewidth]{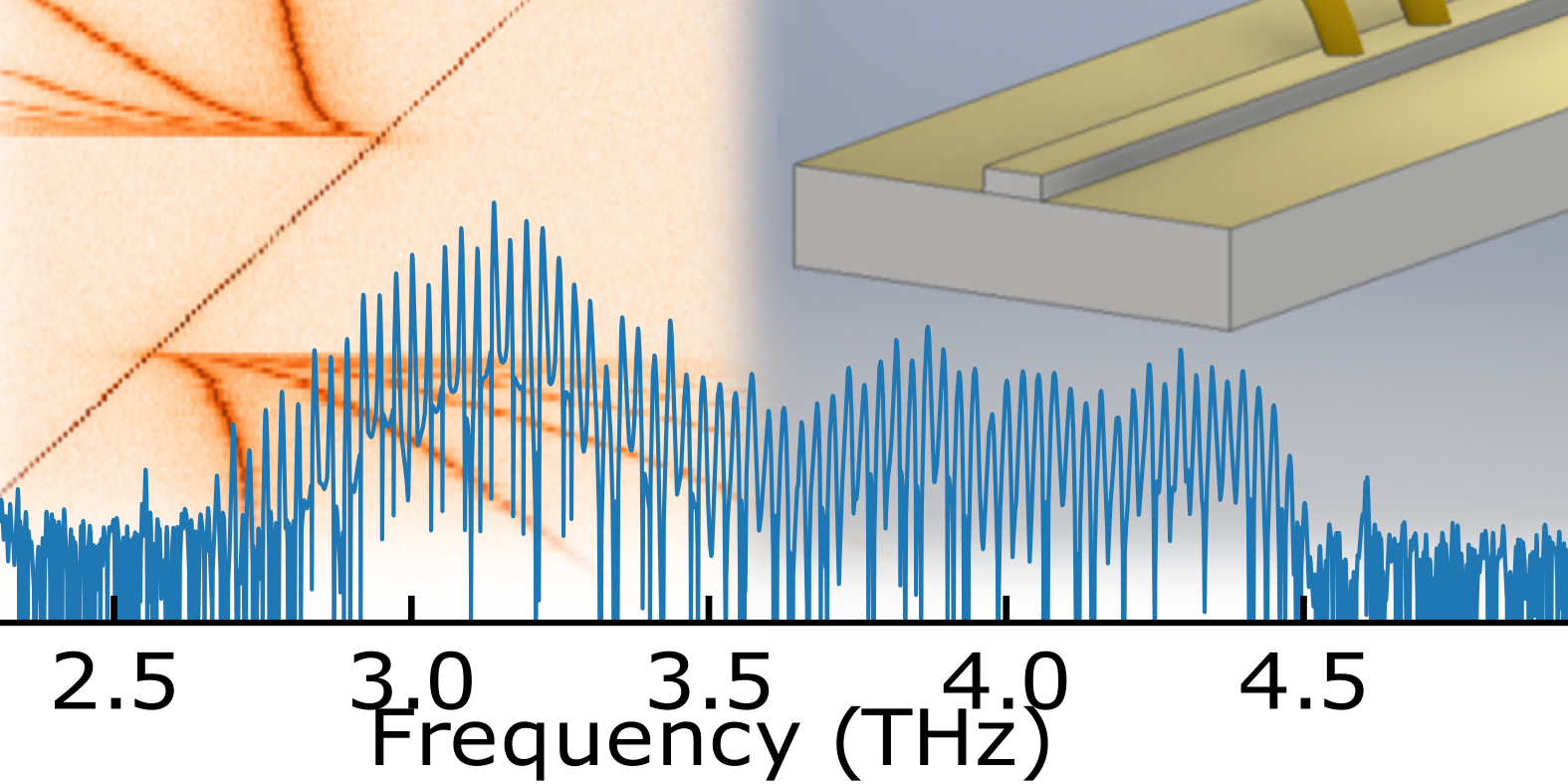}
\newline
Injection Locking map, broad spectrum and sketch of a device. 
\end{tocentry}

\begin{abstract}
  We present homogeneous quantum cascade lasers (QCLs) emitting around 3 THz which display bandwidths up to 950 GHz with a single stable beatnote. Devices are spontaneously operating in a harmonic comb state, and when in a dense mode regime they can be injection locked at the cavity roundtrip frequency with very small RF powers down to -55 dBm. When operated in the electrically unstable region of negative differential resistance, the device displays ultra-broadband operation exceeding 1.83 THz ($\Delta f/f=50\%$) with high phase noise, exhibiting self-sustained, periodic voltage oscillations. The low CW threshold (115 A$\cdot$ cm$^{-2}$) and broadband comb operation ($\Delta f/f=25\%$) make these sources extremely appealing for on-chip frequency comb applications.
\end{abstract}

\section{Introduction}

In recent years, quantum cascade lasers (QCLs) have become increasingly interesting as on-chip sources of mid-IR \cite{Hugi_Nature_2012_MidinfraredFrequencyComb} and THz \cite{Burghoff_Nat.Photonics_2014_TerahertzLaserFrequency} frequency combs, with the main target of spectroscopic applications \cite{Scalari_Appl.Phys.Lett._2019_OnchipMidinfraredTHz, Mottaghizadeh_OpticaOPTICA_2017_5pslongTerahertzPulses, Wang_LaserPhotonicsRev._2017_ShortTerahertzPulse}.
The compact and all-integrated nature of these devices, combined with their high power-per-comb-tooth compared to difference frequency generation (DFG) combs derived from shorter wavelength sources, makes them ideal candidates for portable, high speed dual comb spectrometers \cite{Geiser_BiophysicalJournal_2018_SingleShotMicrosecondResolvedSpectroscopy, Klocke_Anal.Chem._2018_SingleShotSubmicrosecondMidinfrared, Garrasi_ACSPhotonics_2019_HighDynamicRange}. 
The THz devices display several interesting features, as the optical resonator based on a double metal provides a cutoff-free waveguide with a weak frequency depended figure of merit, making these devices well suited for ultra-broadband operation and dual comb spectroscopy \cite{Rosch_Appl.Phys.Lett._2016_OnchipSelfdetectedTerahertz, Yang_Optica_2016_TerahertzMultiheterodyneSpectroscopy, Sterczewski_Opt.Express_2019_ComputationalCoherentAveraging}. Latest developments in high temperature operation of THz QCLs \cite{Bosco_Appl.Phys.Lett._2019_ThermoelectricallyCooledTHz} hold promise also for Peltier cooled operation of fundamental THz combs.

The extreme flexibility and wavelength agility of quantum cascade active regions has allowed for lasing over one octave from the same device \cite{Rosch_Nat.Photonics_2015_OctavespanningSemiconductorLaser}, where different active regions have been combined into the same waveguide. The broadening of the gain spectrum and its shape can be engineered to minimise chromatic dispersion, which is the main limiting factor towards the achievement of a full octave spanning on-chip frequency comb \cite{Rosch_Nat.Photonics_2015_OctavespanningSemiconductorLaser}. However, the heterogeneous cascade concept also comes with some drawbacks; the RF injection locking of these structures is generally more difficult with respect to the homogeneous ones \cite{Li_Opt.ExpressOE_2015_DynamicsUltrabroadbandTerahertz,Gellie_Opt.Express_2010_InjectionlockingTerahertzQuantum} and the design process is prone to uncertainties due to the need for precise matching of the current through the QCL sections.

\begin{figure}
	\centering
	\includegraphics[width=0.48\linewidth]{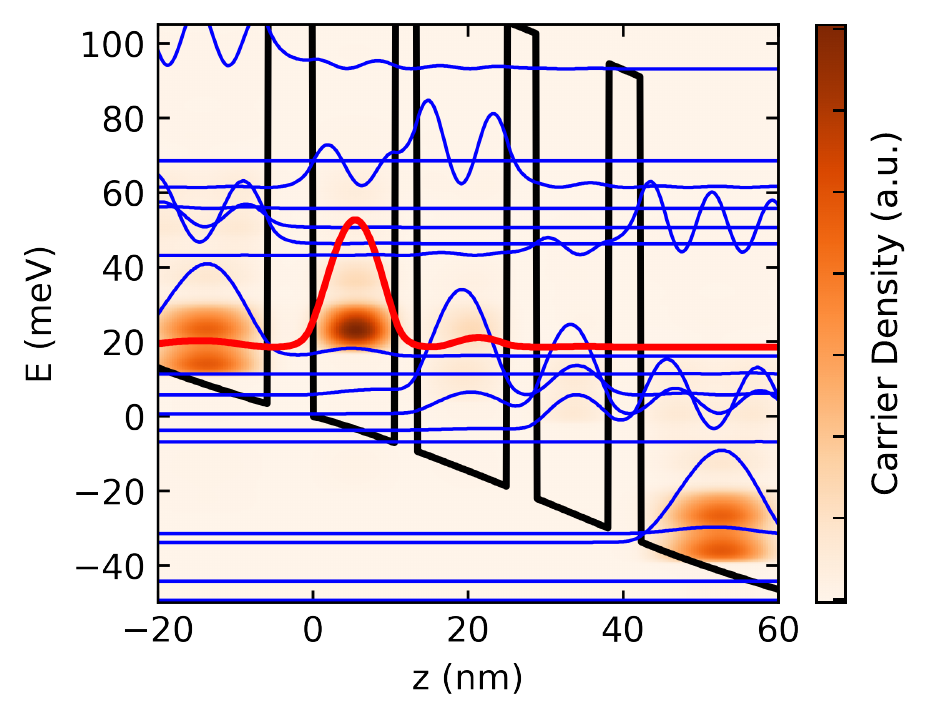}
	\caption{Color scale display of the energetically and spatially resolved electron density computed in the non-equilibrium Green's function model for an applied bias of 50 mV per period and lattice temperature of $T_L = 150$ K. The Wannier-Stark wavefunctions and conduction band potential profile are overlaid. The layer sequence based on GaAs/Al$_{0.15}$Ga$_{0.85}$As is \underline{18.3}/\textbf{5.8}/10.6/\textbf{2.9}/11.5/\textbf{3.9}/9.2/\textbf{4.1} in nm where barriers are in bold and the $2.3\cdot 10^{16}$ cm$^{-3}$ Si doped injection well is underlined. The sequence is repeated 151 times and grown by molecular beam epitaxy (MBE).}
	\label{fig:bandstructure}
\end{figure}

\section{THz QCL design \& operation characteristics}
\label{sec:2}

We present here a QCL design which is derived from our previous 4 quantum well structure \cite{Amanti_NewJ.Phys._2009_BoundtocontinuumTerahertzQuantum}, optimized for wide bandwidth and low threshold operation. The design relies on a high degree of diagonality of the lasing transition.
The lifetime of the upper state and the gain recovery time are also  increased, as already experimentally verified for the previous, less diagonal structure showing a gain recovery time of $\sim$35 ps \cite{Derntl_Appl.Phys.Lett._2018_GainDynamicsHeterogeneousa}, significantly larger than the $\sim$10 ps found for a bound-to-continuum THz design under lasing operation in Ref.~\citen{Markmann_Opt.Express_2017_TwodimensionalCoherentSpectroscopy}. The increased lifetime would in fact change the typically non-resonant optical response to modulation currents for QCLs towards a resonant one, making the design more prone to amplitude modulation up to tens of GHz. \cite{Faist__2013_QuantumCascadeLasersc}

\begin{figure}[htb!]
	\centering
	\includegraphics[width=0.48\linewidth]{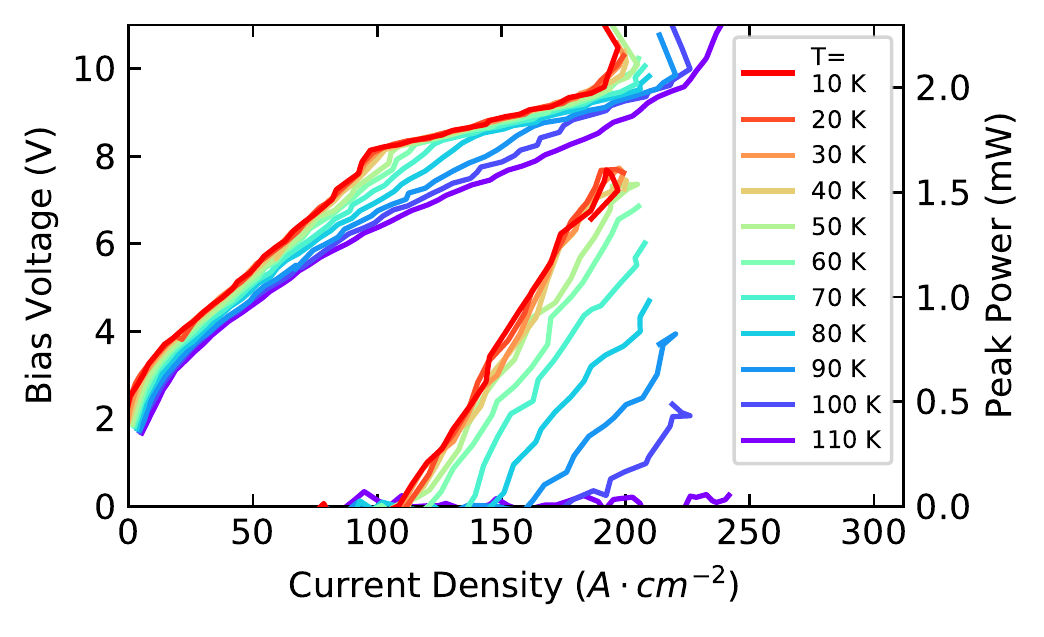}
	\caption{Pulsed LIV curve of a 1.5 mm long, 70 $\mu m$ wide QCL for increasing heat sink temperature. Emission peak power up to 1.6 mW from a single facet and lasing up to 100 K is observed for 1 $\mu s$ long pulses at 100 kHz repetition rate. The output power is uncorrected to any detector size and window losses.}
	\label{fig:LIV}
\end{figure}

The bandstructure of one period of the laser is reported in Fig.~\ref{fig:bandstructure}. The heterostructure was grown by molecular beam epitaxy (MBE) on a semi insulating GaAs substrate in the GaAs/Al$_{0.15}$Ga$_{0.85}$As material system. Thermocompressive Au-Au  (500/500 nm) wafer bonding on a receptor substrate was followed by dry-etching double-metal laser ridges with ICP and top metallization (Ti/Au, 10/200 nm). The ridges feature setbacks on either side for transverse mode control as reported in Ref.~\citen{Bachmann_OpticaOPTICA_2016_ShortPulseGeneration}.   
In Fig.~\ref{fig:LIV} we present the light-current-voltage (LIV) curve as a function of temperature for a 1.5 mm long and 70 $\mu$m wide device operated with a pulse length of 1 $\mu s$ at 100 kHz. For low temperatures we observe an uncorrected single-facet peak output power of 1.6 mW and a threshold current density of 110 A/cm${}^{2}$. The laser operates up to a maximum temperature of 100 K. As expected for strongly diagonal designs both in mid-IR \cite{Faist_Nature_1997_LaserActionTuning} and THz QCLs \cite{Scalari_Opt.ExpressOE_2010_BroadbandTHzLasing} we observe a sharp kink in the IV curve at the laser threshold indicating a pronounced photon driven transport.

\begin{figure}[tb!]
	\centering
 	\includegraphics[width=0.48\linewidth]{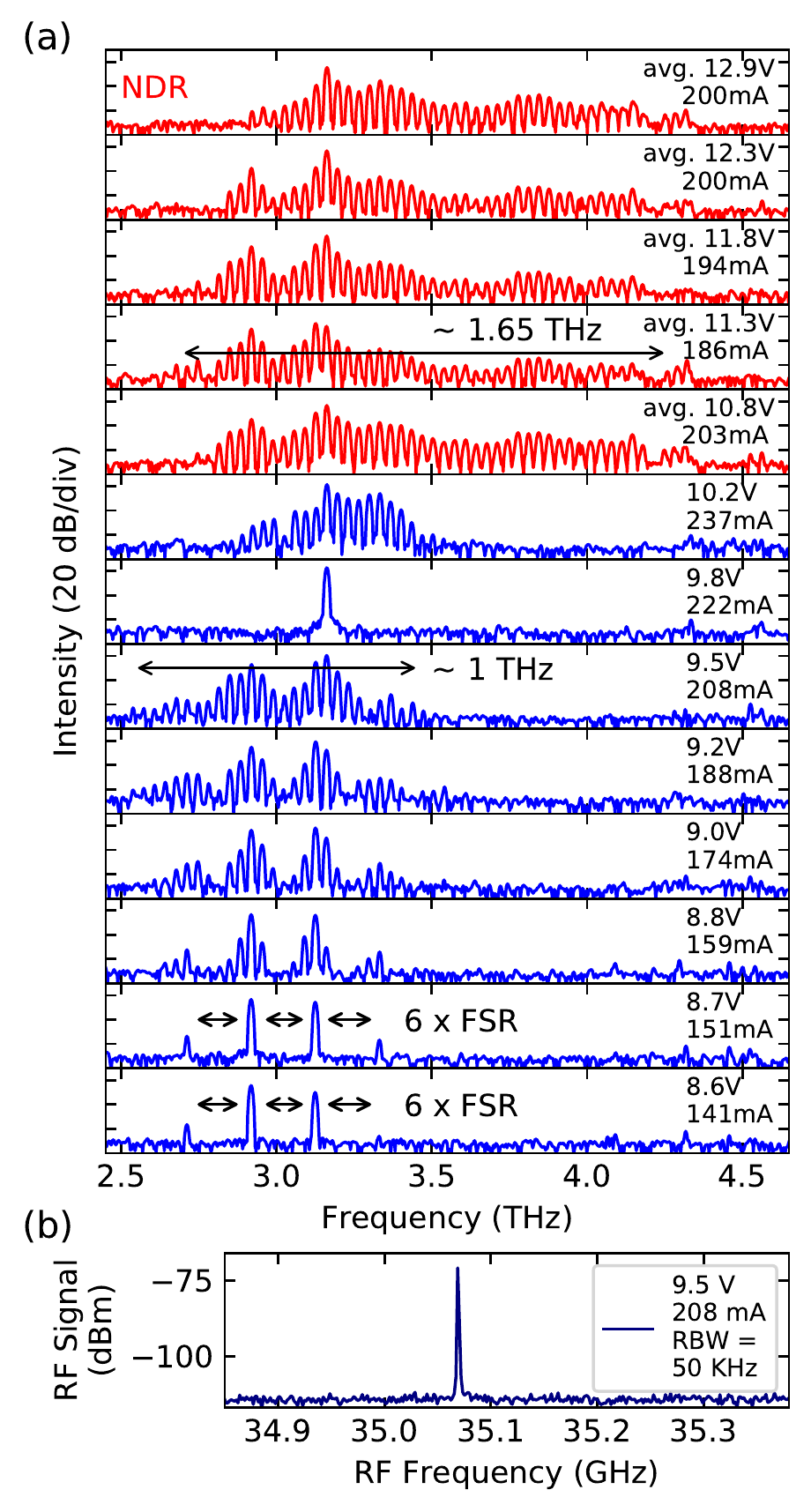}
	\caption{ (a) Evolution of the emission spectrum as a function of the CW applied bias. For low currents we observe self-started harmonics spaced by 6 times the free spectral range (FSR). Increasing voltage leads to a dense spectrum spanning up to nearly 1 THz at 9.5 V, followed by a collapse into a single mode and recovery to dense mode before entering the NDR. The spectra in red are recorded only for voltage driven operation and features emission up to 1.65 THz from this homogeneous QCL. (b) Free-running single and narrow beatnote in the presence of feedback from the FTIR and a S/N of 43 dB at 9.5 V, indicating frequency comb operation.}
	\label{fig:spec}
\end{figure}

\begin{figure}[htb!]
	\centering
	\includegraphics[width=0.48\linewidth]{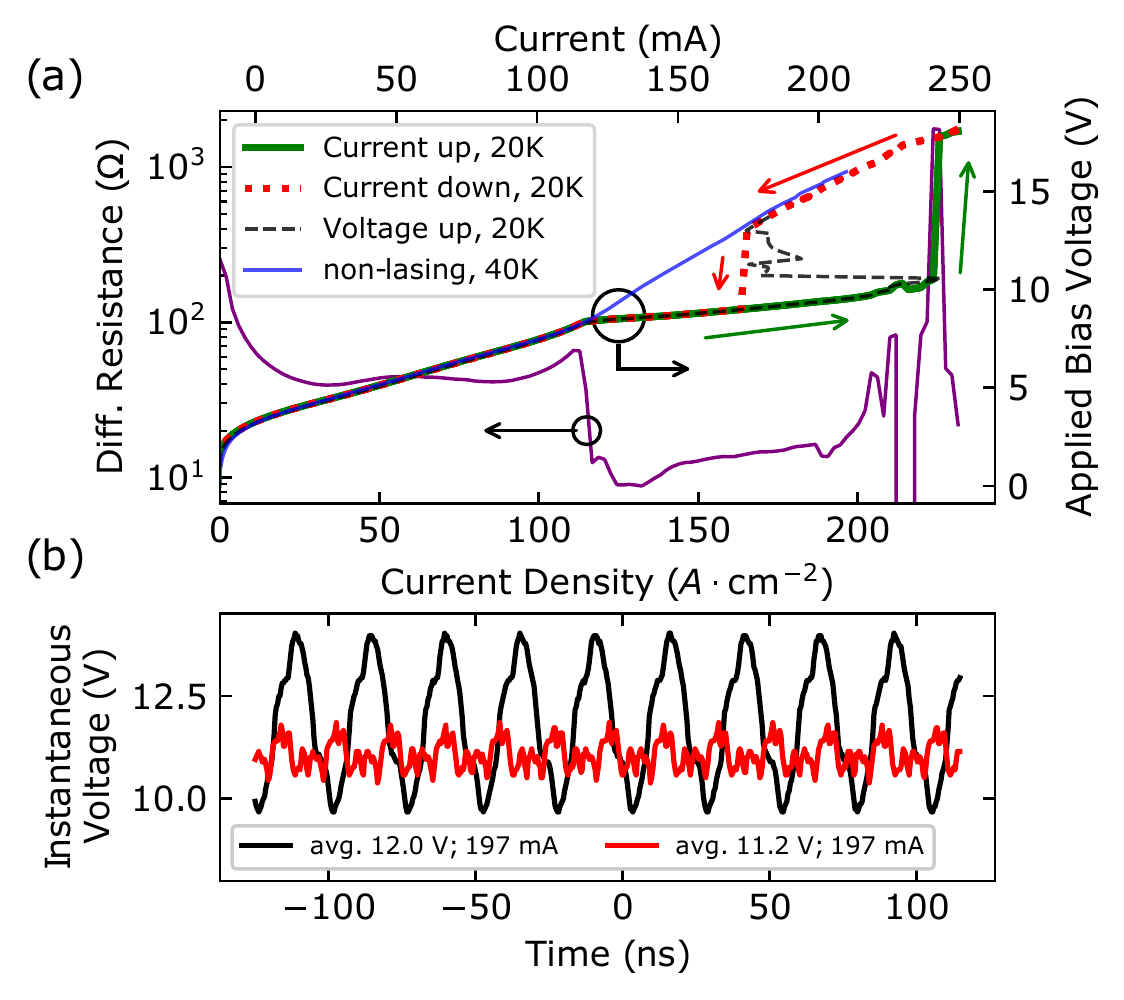}
	\caption{(a): IV curve for increasing current (green), decreasing current (red) and voltage driven (black) operation. For all the pronounced kink in the IV occurs due to the photon driven transport. The current driven modes show a clear hysteresis behaviour without lasing for voltages in the upper branch. The comparison with a non-lasing device (blue) shows that the IV of the lasing device recovers the one of the non-lasing after the NDR. The temperature difference is explained due to the better heat extraction of the smaller non-lasing device. The purple curve shows the differential resistance (of the green curve). (b) Instantaneous voltage over the QCL for two bias points in the voltage driven NDR regime. The two main frequency components are 40 MHz (black) and 54 MHz (red) and are stable over longer time scales.} 
	\label{fig:CWIV}
\end{figure}

In the following, we present a series of measurements which are performed simultaneously, i.e. applied voltage and current in Figs. \ref{fig:spec}, \ref{fig:CWIV} and \ref{fig:BNM} represent the same measurement run to provide the best correlation of the data. The QCL device was mounted in a flow cryostat, aligned to a Fourier transform infrared spectrometer (FTIR) (Bruker, Vertex 80v), powered by a CW current/voltage source (Keithley 2000 with an additional RC filter) over a bias-tee (SHF BT 45D,  1-45 GHz) and connected as well to a spectrum analyzer (SA) (Rohde\&Schwarz, FSW 64) over the bias-tee. An additional bias connection allowed for observing the instantaneous voltage over the QCL with an oscilloscope (LeCroy, HDO6104). The SA and oscilloscope measurements were performed three times before spectrum acquisition and once afterwards for consistency check. 
In Fig.~\ref{fig:spec} (a) we plot a series of CW spectra as a function of applied bias to the laser. Here, the current source was switched to operate as a voltage source allowing to enter the normally inaccessible negative differential resistance (NDR) regime of the QCLs in CW (electronic circuit similar as in Ref.~\citen{Winge_Phys.Rev.A_2018_IgnitionQuantumCascade}). The device is a relatively short Fabry-Perot cavity of 1.2 mm and 90 $\mu$m width.

At low currents in Fig.~\ref{fig:spec} (a), corresponding to the first part of the dashed L-I curve in Fig.~\ref{fig:CWIV} (a) in black, the laser is spontaneously operating on the so-called harmonic state, recently demonstrated and discussed in mid-IR QCLs  \cite{Kazakov_Nat.Photonics_2017_SelfstartingHarmonicFrequency, Piccardo_Opt.ExpressOE_2018_HarmonicStateQuantum}. Such a lasing state is characterised by a spectrum displaying mode separations of several cavity roundtrip frequencies. Its origin lies in the interaction between population pulsations and population gratings in the laser cavity where a $\chi^{(3)}$ nonlinearity is present. We observe harmonic state signatures starting from a bias current of $I=1.22I_\text{thresh}$ and a harmonic mode spacing of $6\times f_\text{rep}=210 $ GHz. To fully prove the coherence of the modes a multi-heterodyne beat against another QCL comb would be necessary. 
After this harmonic state the emission spectrum evolves into a dense multimode state, progressively broadening to span from 2.53 THz to 3.47 THz for $I=208$ mA and 9.5 V. At the same time a single narrow beatnote is observed in Fig.~\ref{fig:spec} (b), indicating comb operation. 
At 215 mA the laser abruptly turns into single mode operation at 3.1 THz: in the IV curve in Fig.~\ref{fig:CWIV} (a) this is visible as a precursor to the strong NDR happening at $I=240$ mA.

Inside the NDR region the instantaneous CW bias voltage is in fact pulsating at RF frequencies, which we have measured by acquiring the bias drop on the laser with an oscilloscope. 

These intensity pulsations group into two different regimes which have main frequency components of 54 MHz resp. 40 MHz and occur for applied voltages of 10.9 V to 11.2 V resp. 11.3 to 12.9 V. For each regime a typical time trace is presented in Fig.~\ref{fig:CWIV} (b) showing a stable intensity pulsation. Similar self-pulsations were already observed in a THz QCL under long current pulse driving close to the laser threshold, and was attributed to domain formation inside the device by Winge \emph{et al.} \cite{Winge_Phys.Rev.A_2018_IgnitionQuantumCascade} supported by a theoretical model (see also Ref.~\citen{Winge_Phys.Rev.A_2018_IgnitionQuantumCascade, Wienold_J.Appl.Phys._2011_NonlinearTransportQuantumcascade, Lu_Phys.Rev.B_2006_FormationElectricfieldDomains, Wacker_Phys.Rep._2002_SemiconductorSuperlatticesModel, Bonilla_Rep.Prog.Phys._2005_NonlinearDynamicsSemiconductor, Esaki_IBMJ.Res.Dev._1970_SuperlatticeNegativeDifferential, Esaki_Phys.Rev.Lett._1974_NewTransportPhenomenon}).  A quantitative view with time resolved measurements is discussed in the next section. In this NDR regime the spectral output of the laser spans an extensive frequency range from 2.6 to 4.3 THz, about 1.7 THz which corresponds to a fractional $\Delta f/f_0=0.5$ with respect to the central frequency $f_0=3.45$ THz. Further devices showed lasing spanning over 1.83 THz in this regime, see Fig.~\ref{fig:Lumi} (a). This represents the highest relative frequency span for a homogeneous quantum cascade laser structure so far.

If the laser structure is driven in current mode, we observe a radically different behaviour;
in Fig.~\ref{fig:CWIV} (a) we plot the overlap of 3 measurements each of the CW current driven IV of a lasing device for increasing (green) resp. decreasing (red) currents at 20 K together with one IV curve of a non-lasing structure (blue) at 40 K.
The current is swept from zero to higher values in the lasing device up to 240 mA (223 A/ cm${}^{2}$, same value as for the voltage driven NDR). At NDR the device shows a hysteresis on the curves of lasing and non-lasing devices. 

By calculating the differential resistance from the voltage driven IV curve in Fig.~\ref{fig:CWIV} (a) we can estimate the ratio $\tau_{eff}/(\tau_{eff}+\tau_2)$ by \cite{Blaser_IEEEJ.QuantumElectron._2001_CharacterizationModelingQuantum, Scalari_LaserPhotonicsRev._2009_THzSubTHzQuantum}
\begin{equation}
\frac{\tau_{eff}}{\tau_{eff}+\tau_2} = 1-\frac{R_{d,S>0}}{R_{d,S=0}} ,
\end{equation}
where $R_{d,S>0}$ resp.~$R_{d,S=0}$ is the differential resistance above resp. below threshold, $\tau_2$ the lower state lifetime and ${\tau_{eff} = \tau_3\cdot (1-\tau_2/\tau_{32})}$ with $\tau_3$ the upper state lifetime and $\tau_{32}$ the non-radiative lifetime. Entering the values from Fig.~\ref{fig:CWIV} (a) we find $\tau_{eff}/(\tau_{eff}+\tau_2) \approx 0.81$. This value is roughly twice that reported in Ref.~\citen{Scalari_LaserPhotonicsRev._2009_THzSubTHzQuantum} and indicates a long upper state lifetime. This result is also supported by the non-equilibrium Green's function (NEGF) simulations in the following section, where the corresponding scattering rates calculated in the Wannier basis provides a ratio of 0.83 at 50 mV/period and $T_L$ = 150 K. Additionally, we find that the diagonal transition has a relatively low computed maximum oscillator strength between the upper state and the lower states of $\sum_{low_j} f_{up,~low_j} \approx 0.25-0.20$ from the lasing threshold to the NDR.

\section{Non-Equilibrium Green's Function Simulation and Time-Resolved Spectrum}

\begin{figure}[htb!]
	\centering
	\includegraphics[width=0.48\linewidth]{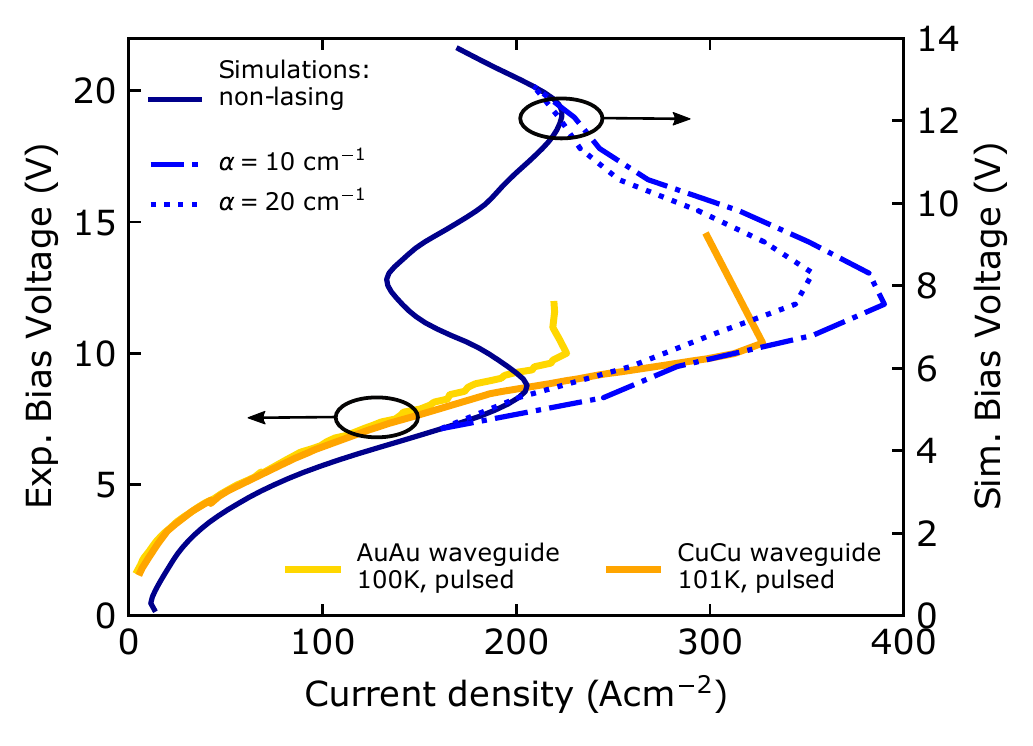}
	\caption{NEGF simulated LIV at 150 K (lattice temperature, dark blue dashed), NEGF including photon driven transport clamped to losses of 10 cm$^{-1}$ resp. 20 cm$^{-1}$. Experimental LIV curves for gold and copper devices at ~100 K (heat sink temperature) are presented. The voltage difference of simulation to experiment is attributed to the Schottky contact and additional serial resistances in the real device.}
	\label{fig:LIV_Sim_Exp}
\end{figure}

In order to explain the experimental observations of a pulsating laser in the NDR region, we now present NEGF~\cite{Wacker_IEEEJ.Sel.Top.QuantumElectron._2013_NonequilibriumGreenFunction} simulations of the laser under operation, as well as time-resolved spectrum measurements. In the NEGF formalism used here, the photo-induced current can be included by increasing the ac field strength in the simulations until the gain clamps at some assumed value for the total losses \cite{Lindskog_Appl.Phys.Lett._2014_ComparativeAnalysisQuantum}. At a lattice temperature of $T_L=150$ K there is still enough gain for significant photo-driven transport, even for losses as high as 10 cm${}^{-1}$ and 20 cm${}^{-1}$, as shown in Fig.~\ref{fig:LIV_Sim_Exp}. This performance is most likely overestimated since electron-electron scattering has been neglected in these simulations, which can significantly deteriorate the performance \cite{Franckie_Appl.Phys.Lett._2018_TwowellQuantumCascade}. The simulations show that the laser turns on at the first alignment peak close to 6 V, and the strong photo-induced current allows the laser to operate into the bias region where the non-lasing device shows NDR (indicated by the dark-blue dashed line). Such a clear NDR feature is not seen in the experimental non-lasing device driven in voltage mode, most likely because of leakage currents through high lying parasitic states. At a bias around 8 V the gain begins to drop and as a result so does the photo-driven current, which creates an NDR region for the lasing device where in the absence of lasing the differential resistance returns to positive values. Above this bias, the laser operates with high optical power in an unstable electrical region. This can explain the 
broad spectra observed in this bias region for the experimental device in Fig.~\ref{fig:spec}, as the laser can alternate between different bias points at a constant current density. In addition, electric field domains can be formed where separate sections of the laser are biased with different voltages simultaneously.

Since the peak current density shows a strong temperature dependence a quantitative comparison between the simulated currents at 150 K with the LIV at low temperatures for the device employing Au-Au waveguide is not possible. Therefore, we also fabricated a device with a Cu-Cu waveguide which showed significant photo-driven transport even at 100 K cryostat temperature, as seen by the orange solid line in Fig.~\ref{fig:LIV_Sim_Exp}. Just as in the simulations, at the alignment bias where the slope of the IV curve flattens out, the device starts to lase and pushes the operation opposite to the non-lasing NDR region. At a higher bias, the current starts to again decrease as the gain diminishes. The difference in bias between the experimental and simulated curves is explained by an estimated contact Schottky bias drop of 0.9 V for the Au-Au device and 0.6 V for the Cu-Cu device, and an additional series resistance providing lower slope in the IV curve.

\begin{figure}[htb!]
	\centering
	\includegraphics[width=0.48\linewidth]{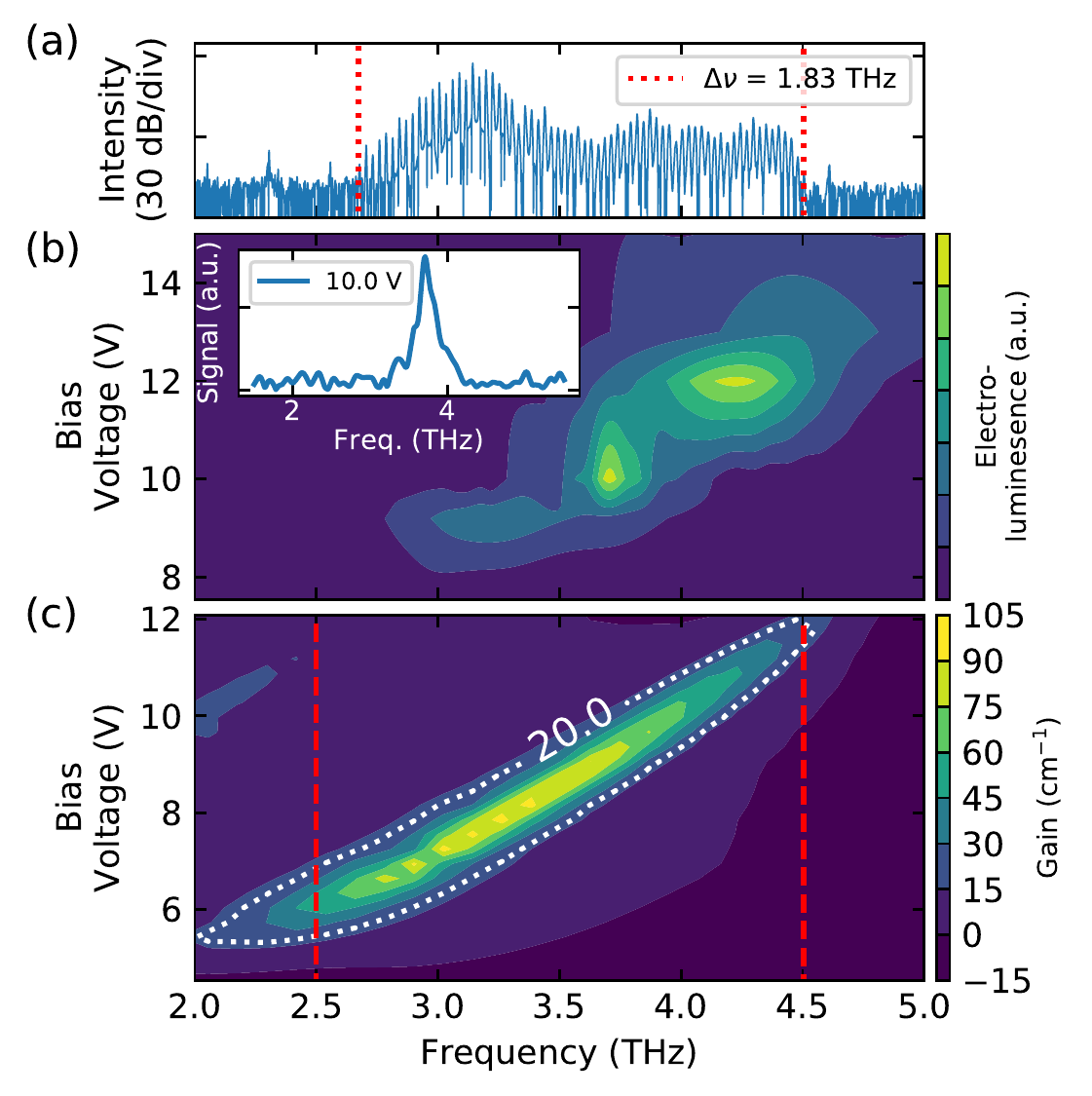}
	\caption{(a) 1.83 THz broad spectrum of a 1.45 mm long device operating in CW. (b) Electroluminesence measurement of a non-lasing device with respect to the simulated gain by NEGF in (c). The lasing spectral region, taking the minimum and maximum frequency at all measured bias points, is indicated by the red dashed lines. The emission in (a) covers 90 \% of NEGF predicated lasing frequencies. Difference in bias voltage are explained by additional serial resistance in the experiment like the Schottky contact of the laser.}
	\label{fig:Lumi}
\end{figure}

The LIV curve provides insights on the electrical origin of the self-pulsation, but additional gain simulations are necessary to explain the broadened spectrum. Also more sophisticated models rely on accurate gain predictions as explained in Ref. \cite{Winge_Phys.Rev.A_2018_IgnitionQuantumCascade}.
Therefore, we now compare the simulated and experimental emission spectrum evolution in Fig.~\ref{fig:Lumi}. Panel (a) presents the broadest spectrum spanning 1.83 THz of a 1.45 mm long device in the intensity pulsating regime with high phase noise. The spectrum is compared additionally to the electroluminescence measurement of a non-lasing device (short cavity of 0.345 mm) in Fig.~\ref{fig:Lumi} (b) and to the simulated gain in Fig.~\ref{fig:Lumi} (c). The bandwidths of the electroluminesence and the simulated gain agree reasonably with each other and as well with the minimal and maximal observed frequency in all measured devices indicated by the red dashed lines in (c). However, the bandwidth at a given bias point is significantly narrower than the spectrum in (a). In the simulation the 20 cm${}^{-1}$ contour line is marked to represent the estimated value of the waveguide losses. The simulated gain curve reproduces well the trend of frequency vs.~bias, although the simulations predict emission at frequencies below 2.5 THz. Actually, the waveguide losses are non-uniform and increase for lower frequencies due to the presence of lossy side-absorbers and can explain the absence of lower modes in the experiments. In the inset of Fig.~\ref{fig:Lumi} (b) we plot an electroluminescence spectrum taken at 10 V bias where a $\sim$1 THz broad signal is measured agreeing with the $\sim$1 THz spectra observed in Fig.~\ref{fig:spec}.

Connecting to the above discussion regarding laser operation in the photo-driven NDR region, combining the simulated spectra around 7 V with those around 10 V would reproduce a spectral width similar to that observed in Fig.~\ref{fig:Lumi} (a). A photo-driven domain formation could therefore explain the observed broad spectra in the high phase-noise regime. However, such a phenomenon would require a more sophisticated modeling as the different domains would affect the current-voltage relationship in an intricate manner.

\begin{figure}[htb!]
	\centering
	\includegraphics[width=0.48\linewidth]{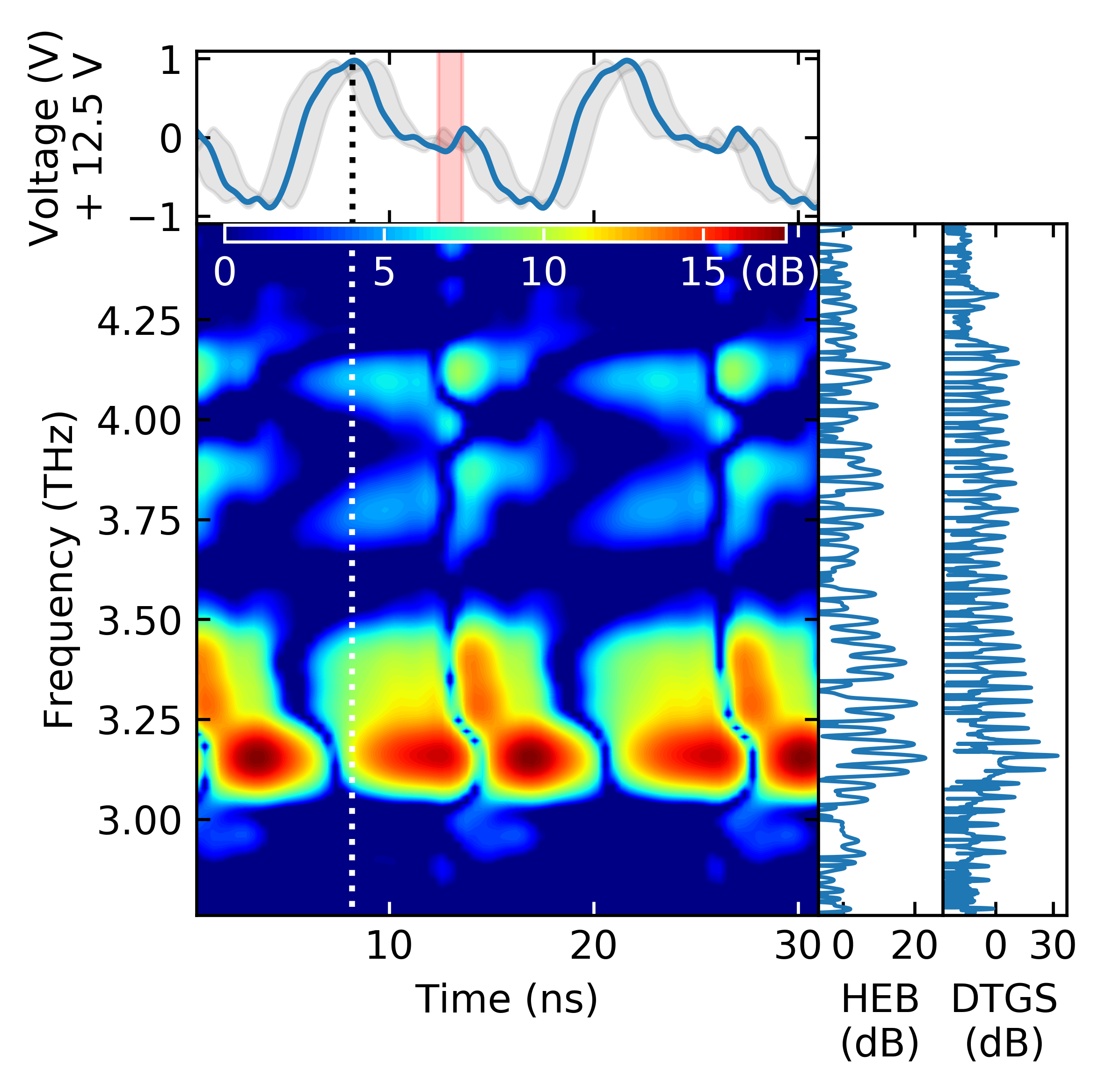}
	\caption{Top: Stable voltage oscillation over the device bias. Red shaded indicates the time interval used for getting the mean interferogram point from the detector signal. Grey shaded indicates the delay uncertainty between voltage signal and spectrogram. Right: Spectra recorded by the internal DTGS and external, fast superconducting HEB detector. Colormap: Spectrogram of shortened interferograms for visibility reasons. An animation of the time-resolve data is provided in the supplementary using the full interferogram length. Clear changes resp. oscillations of the different emission modes are observed and qualitatively support the argumentation of oscillating between multiple solution for constant current when entering the NDR regime and can be seen as oscillation between the ~7 V and ~10 V bias point in the NEGF simulation as can be seen in  Fig.~\ref{fig:LIV_Sim_Exp},\ref{fig:Lumi} } 
	\label{fig:DomainOsc}
\end{figure}

Nevertheless, spectral pulsation of the device can be directly measured due to the self-sustained pulsation of the bias voltage. Therefore, we perform a step-scan FTIR measurement with a fast detector (Superconducting Hot-Electron Bolometer (HEB) by \textit{Scontel}) and trigger on the FTIR step followed by a trigger on the self-sustained oscillation with a fast oscilloscope. A 2 $\mu$s long time trace of the detector and voltage signal is recorded and by post-processing sliced into the oscillation period length and averaged. A $\sim$20 ns time delay due to different travelling times for the voltage signal (cables) and detector signal (optical path + cables) is measured (cables) resp.~calculated (optical path) and accounted for in the analysis. In the end we can reconstruct an interferogram for each point on the voltage oscillation, i.e as a function of time. Due to the detector's $\sim$200 MHz amplifier bandwidth limit and signal-to-noise consideration the data is further averaged over 1.2 ns time intervals. 
The resulting spectrogram is shown in Fig.~\ref{fig:DomainOsc} together with the voltage signal (top) and the spectra from the total HEB signal and the reference measurement of the FTIR deuterated triglycine sulfate (DTGS) detector (right). The gray shaded area around the voltage oscillation indicates the uncertainty in the delay between voltage signal and spectrogram, whereas the red shaded region shows the 1.2 ns window for averaging. For visibility short interferograms are used for the spectrogram, whereas for the HEB spectrum (vertical side of Figure) the full interferogram length was taken. Therefore, the spectrogram shows the spectral regions which are present for each time step rather than the individual modes. An animation of the time resolved spectra, taking the hole interferogram, is provided in the supplementary material. From the spectrogram we deduce that there are clear periodic changes in the emission spectrum during one period of the voltage oscillation. These results suggest domain formation occuring inside the QCL, rather than the bias changing uniformly over the whole structure. 

\section{Injection locking \& Coherence properties}

\begin{figure}[htb!]
	\centering
	\includegraphics[width=0.48\linewidth]{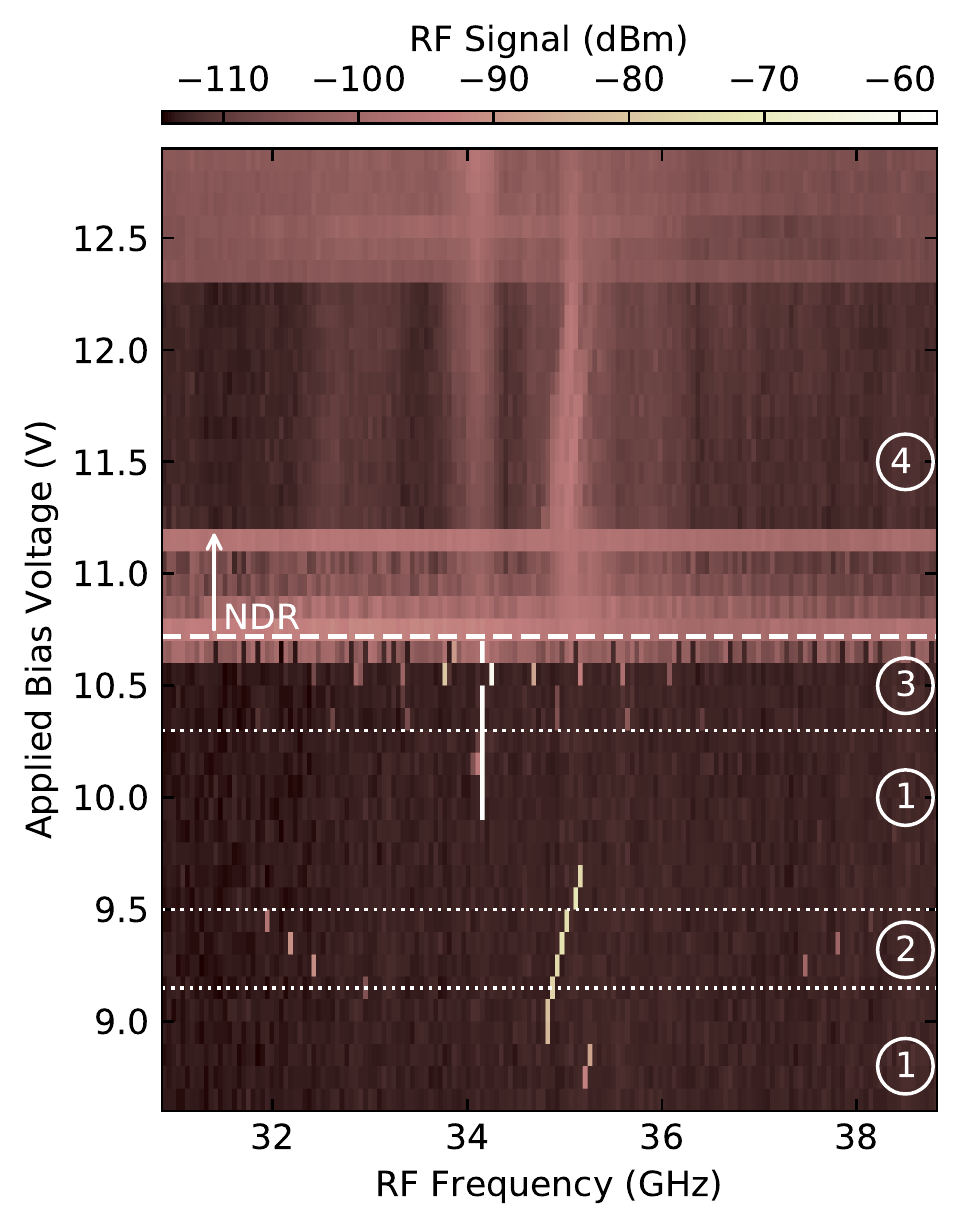}
	\caption{Simultaneously recorded beatnote map to the spectra in Fig.~\ref{fig:spec} (a). The RF axis is binned of 30 values for better visibility. In regions (1) a single and narrow beatnote with a S/N up to 43 dB is observed. Region (2) resp. (3) show a single narrow beatnote with sidebands. In region (4), in the NDR, a broad phase noise is observed. Still there are two clear regions corresponding to the stable beatnotes below NDR.} 
	\label{fig:BNM}
\end{figure}

We now discuss the spectral coherence properties of the studied laser as a function of the different biasing regimes.
In Fig.~\ref{fig:BNM} we present a beatnote map recorded with a spectrum analyser which extracts the signal with a bias-tee from the bias line of the laser. The beatnote map was recorded simultaneously to the spectra in Fig.~\ref{fig:spec} (a) and the data is binned along the RF frequency axis to make the beatnote clearly visible in the colormap.
Below the NDR there are regions where one (label (1)) or several extremely narrow beatnotes (label (2), (3)) appear. In the regions labeled as (2) resp. (3) the multi-mode appearance is actually sidebands of the main narrow beatnote \cite{Li_Opt.ExpressOE_2015_DynamicsUltrabroadbandTerahertz} which are amplitude resp. frequency modulated. When the laser enters the NDR region the beatnote broadens, showing that the laser is operating in a high phase noise regime labeled as (4). Here, two broad signals can be observed around 34 GHz and 35 GHz, respectively. Each of these two signals have their corresponding narrow beatnote in the stable regimes (1) and (3) for the signal around 34 GHz, and regimes (1) and (2) for the signal around 35 GHz. This suggests that the broad spectrum observed in the NDR region is due to both the low-bias and high-bias regimes being present at the same time in the QCL.

\begin{figure}[htb!]
	\centering
	\includegraphics[width=0.48\linewidth]{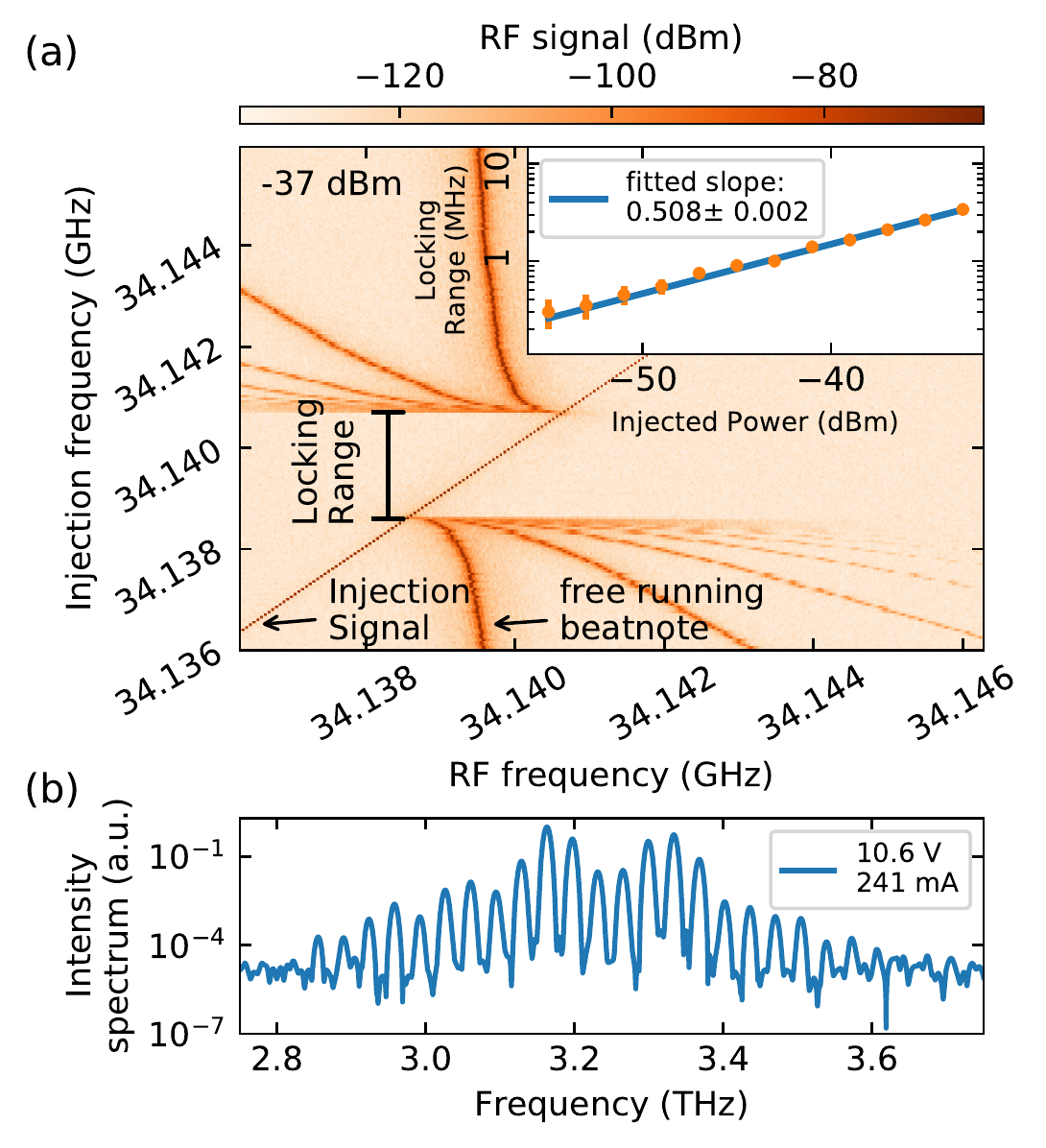}
	\caption{(a) Injection locking map for -37 dBm at the QCL. The injection signal is swept pulling the free running beatnote towards itself and essentially locking it. The typical sidebands are clearly visible. Inset: Locking range for different injection powers following the square-root dependence of the Adler's equation (slope of 0.5 in log-log scale). The fit was weighted by the locking ranges. (b) Recorded spectra at 241 mA, 10.6 V and 20 K with injection-locked beatnote with power of -26 dBm at the device. The spectrum is similar to the free running beatnote spectrum.}
	\label{fig:inj}
\end{figure}

Providing the bias line with another bias-tee and employing a microwave generator we can investigate the frequency locking characteristics of the laser structures in the stable operating regime \cite{Hillbrand_NaturePhoton_2019_CoherentInjectionLocking}. The cryostat is equipped with 40 GHz RF cables and 18 GHz SMA connectors at the cryo finger connected to gold pads followed by gold wires to the QCL. The total losses from cables, bias-tee, connectors, gold pads and wires are estimated to be 25 dB around 35 GHz, deduced from transmission measurements. A typical injection locking map is reported in Fig.~\ref{fig:inj} (a) for an estimated injected microwave power of -37 dBm at the QCL. It is evident that the laser beatnote is pulled towards the injected signal and successively locked. The sidebands, which are generated, are typical for this kind of experiments \cite{St-Jean_LaserPhotonicsRev._2014_InjectionLockingMidinfrared}. A maximum locking range of 4 MHz is achieved for an injected power of -33 dBm, where we injection lock more than 20 laser modes. The reported locking range as a function of the injected microwave power in the inset of Fig.~\ref{fig:inj} (a) shows a very good agreement with  Adler's model equation which describes the injection locking as a function of the Injected power, the RF quality factor of the slave resonator and the injection frequency \cite{Siegman__1986_Lasersb}. Our results compare well with the ones reported in Ref.\cite{Gellie_Opt.Express_2010_InjectionlockingTerahertzQuantum} where a THz QCL operating in double metal waveguide with a beatnote at very similar frequency was investigated. Our injection locking range is a factor of 10 smaller, and this could be ascribed to the non-optimal RF matching of our cryo head which translates into an effectively higher quality factor for the RF cavity. In Fig.~\ref{fig:inj} (b) we measured the spectrum of the lasing device at 241 mA, 10.6 V at 20 K under injection-locking with -26 dBm power at the laser. For these injection powers the emission spectra typically stayed unchanged.

\begin{figure}[htb!]
	\centering
	\includegraphics[width=0.48\linewidth]{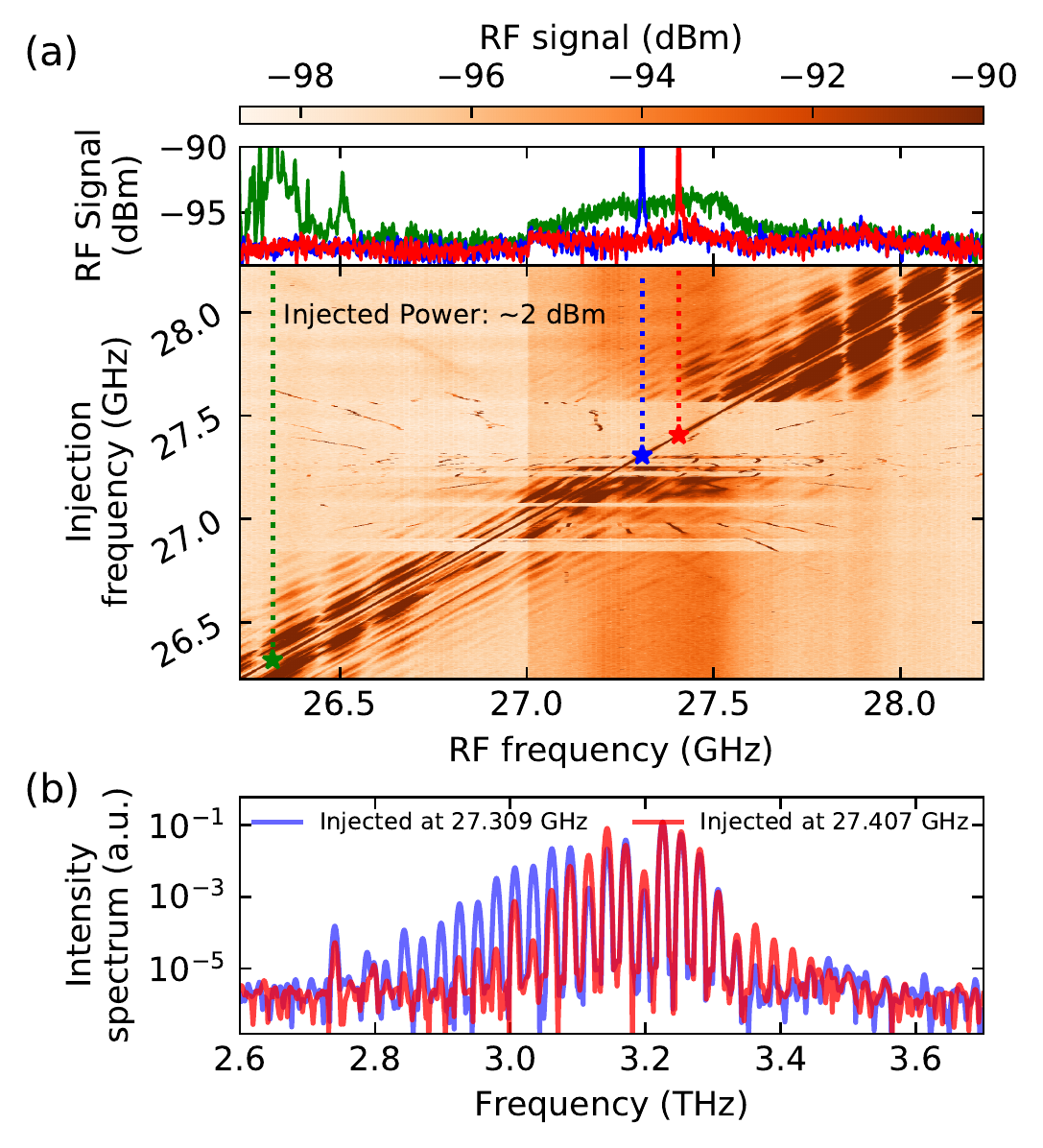}
	\caption{(a) High phase noise regime below NDR of a 1.45 mm long device. The injection map is binned to groups of 5 along RF frequency axis and signal above -90 dBm is cut for visibility. The change at 27 GHz is due to the spectrum analyser. By sweeping the injection frequency we observe locking of the modes, single curves in blue and red without any side band, indicated by reduction of the high phase noise signal in green to the noise floor. (b) Spectra at the two injection points marked in (a) showing that the spectral shape is changed significantly depending on the injection frequency.}
	\label{fig:injHPN}
\end{figure}

Furthermore, we investigate injection locking in the high phase noise regime, i.e. with a broad RF signal. Such a regime below the NDR is presented in Fig.~\ref{fig:injHPN} (a) in green, where the injection is still far from the RF signal. The injection power at the laser is estimated to be 2 dBm, and the signal is cut above -90 dBm and binned into groups of 5 along the RF frequency axis for visibility. The modulation of the injection sidebands along the RF frequency axis is due to the resonances of the circulator used . The abrupt kink at 27 GHz is an artifact from the spectrum analyzer. The high phase noise signal has an SNR of roughly 3 dBm and spans from 27 GHz to 27.6 GHz. By sweeping the injection frequency we observe multiple regions where the high phase noise signal can be locked or partially locked, i.e presence of side bands, to the reference signal. In these regions the noise floor is recovered as presented in Fig.~\ref{fig:injHPN} (a) in blue and red indicating locking of the modes. In the regime of locked beatnote we recorded the spectra, see Fig.~\ref{fig:injHPN} (b). We are clearly able to substantially change the emission spectrum depending on the injection frequency.  

\begin{figure}[htb!]
	\centering
	\includegraphics[width=0.48\linewidth]{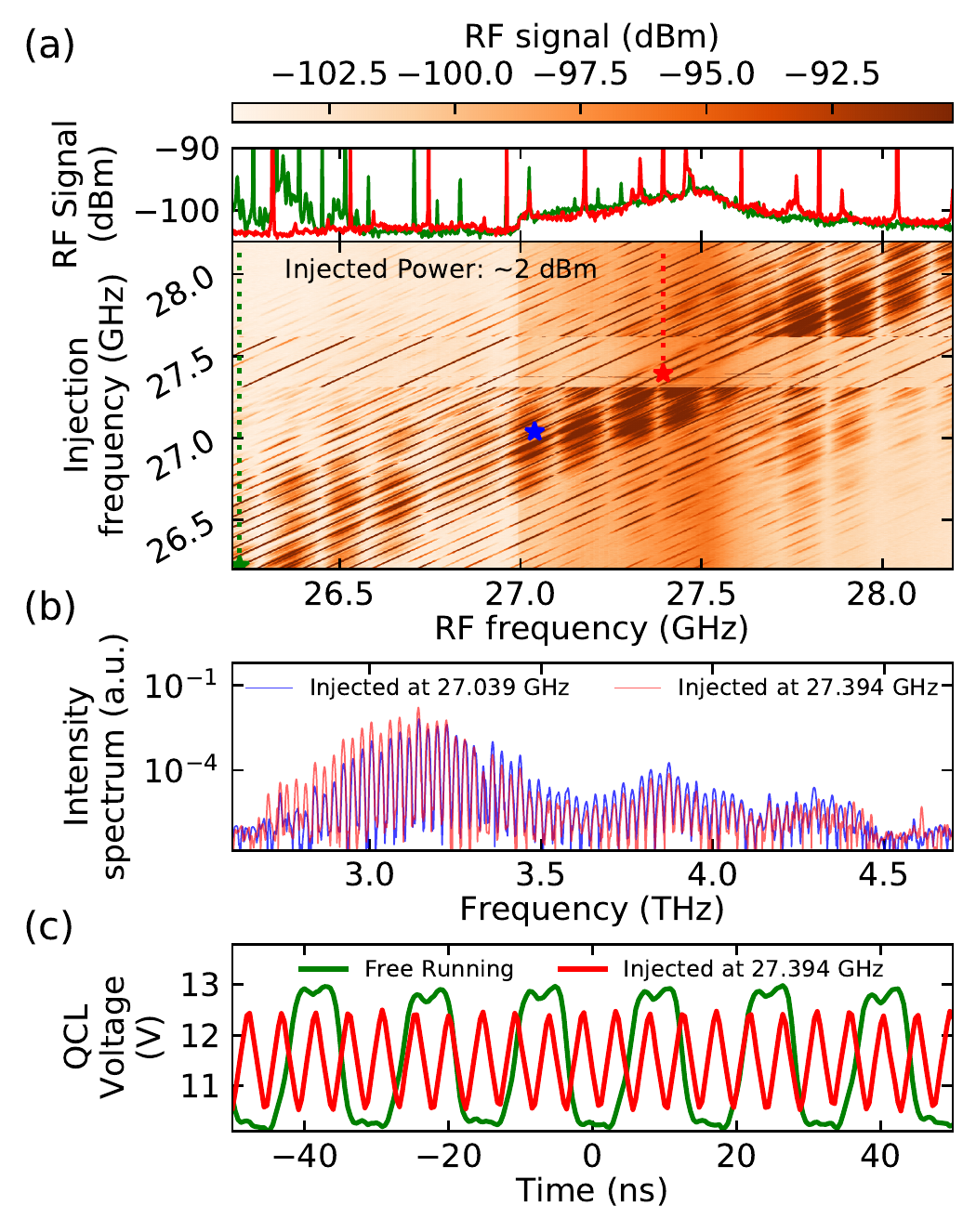}
	\caption{(a) High phase noise regime in the NDR of a 1.45 mm long device. The injection map is binned to groups of 5 along RF frequency axis and signal above -90 dBm is cut for visibility. The change at 27 GHz is again due to the spectrum analyser. By sweeping the the injection the change of side bands is observed. The high phase noise regime stays and no locking is observed. (b) Spectra at the two injection points marked in (a) showing that even though the device is not locked, the spectral shape is changed. (c) The voltage oscillation of the free running in green and injected in red are displayed and match the corresponding side band spacing in (a).}
	\label{fig:injHPNNDR}
\end{figure}

By switching the operation point to the NDR, i.e. having voltage oscillations, we repeat the previous injection locking experiment. The results are shown in Fig.~\ref{fig:injHPNNDR}.  There are two main differences compared to the former: First, the injection signal shows many side modes and in the regime from 27.35 GHz to 27.65 GHz we observe a change in the side bands with keeping the high phase noise nearly unchanged. The side bands arise from the QCL voltage oscillation shown in Fig.~\ref{fig:injHPNNDR} (c),  green curve, whose main frequency component is 63 MHz, fitting to the side band spacing. Sweeping the injection frequency further, the side band spacing changes to 216 MHz, corresponding to new a voltage oscillation (red curve of Fig.~\ref{fig:injHPNNDR}~(c)). At this point the modes are not locked to the injection, as can be seen from the unchanged high phase noise. Comparing the spectra of the two oscillating regions we observe again a change in the spectra due to the injection. Interestingly the changed oscillation \textit{remains stable} when switching off the injection signal, even in the presence of FTIR feedback. Hence, multiple stable solutions of the voltage oscillation and consequently of the domain formation exist for the same bias point.

\section{Conclusion}
We have presented a broadband, homogeneous THz QCL with a low current threshold. The emission bandwidth is roughly 1 THz with a simultaneously narrow and strong beatnote, indicating comb operation. By entering the NDR regime, lasing up to a bandwidth of nearly 1.83 THz was observed. A reasonable agreement of measured LIV and spectral width of the devices with NEGF simulation was found, which together with the time-resolved spectral measurements suggest domain formation inside the device as the origin of the observed wide emission spectrum and self-pulsations. Finally, the low power injection locking capability of the device was demonstrated and by a strong RF injection in the high phase noise regime we observed a clear influence on both the laser emission and the self-pulsating behaviour. Therefore, the presented active region design can serve as a building block towards broadband, on-chip, octave-spanning THz QCL combs\cite{li2019compact}.

\begin{acknowledgement}

We acknowledge the financial support from H2020 European Research Council Consolidator Grant (724344) (CHIC) and (766719) (FLASH) and from Schweizerischer Nationalfonds zur F\"orderung der Wissenschaftlichen Forschung (200020-165639) as well as the help form P. T\"aschler and discussions with C. Sirtori and S. Barbieri.

\end{acknowledgement}

\begin{suppinfo}

Supporting Information Available:
\begin{itemize}
  \item Time-Resolved-Spectrum.mp4: The movie shows the spectral evolution in time with respect to the observed bias voltage oscillation. The data set is the same as for Fig.7 in the Manuscript but full length taken. It provides an alternative view on the measurement results.
\end{itemize}

This Material is available free of charge via Internet at http://pubs.acs.org

\end{suppinfo}

\providecommand{\latin}[1]{#1}
\makeatletter
\providecommand{\doi}
  {\begingroup\let\do\@makeother\dospecials
  \catcode`\{=1 \catcode`\}=2 \doi@aux}
\providecommand{\doi@aux}[1]{\endgroup\texttt{#1}}
\makeatother
\providecommand*\mcitethebibliography{\thebibliography}
\csname @ifundefined\endcsname{endmcitethebibliography}
  {\let\endmcitethebibliography\endthebibliography}{}

\end{document}